\documentclass[a4paper,10pt]{article}
\usepackage[utf8]{inputenc}
\usepackage{url}
\usepackage{xcolor}
\usepackage[a4paper, left=1.5cm, right=1.5cm, top=2cm, bottom=2.5cm]{geometry}
\usepackage[linkcolor=blue, citecolor=blue, urlcolor=blue, colorlinks=true]{hyperref}
\usepackage{multicol}
\usepackage{graphicx}
\usepackage[affil-it]{authblk}
\usepackage{hyperref}
\usepackage{titlesec}
\usepackage{float}

\titleformat*{\section}{\large\bfseries}

\titleformat*{\subsection}{\normalsize\itshape}
\titleformat*{\subsubsection}{\small\bfseries}
\titleformat*{\paragraph}{\large\bfseries}
\titleformat*{\subparagraph}{\large\bfseries}

\title{\textbf{An algorithm for onset detection of linguistic segments in continuous electroencephalogram signals}}
\author{\normalsize Tonatiuh Hernández-Del-Toro\thanks{Email: \textbf{tonahdztoro@gmail.com}} }
\author{\normalsize Carlos A. Reyes-García}
\affil{\small Biosignal Processing and Medical Computing Lab. Instituto Nacional de Astrofísica, Óptica y Electrónica. Mexico.}

\date{}

\begin{document}

\maketitle

\noindent\makebox[\linewidth]{\rule{\textwidth}{0.4pt}}
\textbf{Abstract}

\noindent A Brain Computer Interface based on imagined words can decode the word a subject is thinking on through brain signals to control an external device. In order to build a fully asynchronous Brain Computer Interface based on imagined words in electroencephalogram signals as source, we need to solve the problem of detecting the onset of the imagined words. Although there has been some research in this field, the problem has not been fully solved. In this paper we present an approach to solve this problem by using values from statistics, information theory and chaos theory as features to correctly identify the onset of imagined words in a continuous signal. On detecting the onsets of imagined words, the highest True Positive Rate achieved by our approach was obtained using features based on the generalized Hurst exponent, this True Positive Rate was 0.69 and 0.77 with a timing error tolerance region of 3 and 4 seconds respectively.

\noindent \textit{Keywords:} Imagined Speech, Onset Detection, Continuous Signal.\\
\noindent\makebox[\linewidth]{\rule{\textwidth}{0.4pt}}

\begin{multicols}{2}

%%%%%%%%%%%%%%%%%%%%%%%%%%%%%%%%%%%%%%%%%%%%%%%%%%%%%%%%%%%%%%%%%%%%%%%%%%%%%%%%
\section{Introduction}
\hspace{\parindent} A Brain Computer Interface (BCI) based on imagined speech (hearing self-voice internally without any muscular movement), can decode the syllable, vowel or word a subject is thinking on through brain signals in order to control an external device, this technology can be used to improve life conditions of persons with disabilities and to improve our natural human abilities in many technological fields.

Imagined speech as electrophysiological source has shown very promising results in recent years, however, there are still some problems that are not fully solved \cite{Song2017a}. Imagined speech can be exploited in many ways (syllable, vowel or word). Among all these, we will focus in imagined words.

In the task of building a fully asynchronous BCI that takes electroencephalogram (EEG) signals as input and uses imagined words as electrophysiological source, many problems need to be solved, among them, is the problem of detecting the onset of imagined words in a continuous signal, this is, to correctly identify when the user starts to imagine a word. This requires to correctly classify between mental linguistic activity and idle states. This problem, if solved, can lead to a BCI based on imagined words that is activated in the exact moment the user desires it to.

To the best of our knowledge, there has not been an approach to solve the problem of identifying the onset of imagined words in continuous EEG signals. However, in \cite{Song2017} they tried to identify the onset of high pitch sound imagery production.

The present work describes an algorithm to detect the onset of imagined words in a continuous EEG signal using common statistical values, the Shannon entropy and the generalized Hurst exponent as features. The dataset used in this work is described in \cite{Torres-Garcia2016}, which has recordings of 27 subjects imagining 5 different words in Spanish needed to control a PC pointer.

The algorithm proposed uses linguistic and nonlinguistic segments to train a classifier for latter sequential evaluation on a continuous signal to predict where the onsets of imagined words are. The measure of performance is the True Positive Rate (TPR). This is, the number of onsets that are correctly identified divided by the total number of onsets.

For the feature set based on statistical values, our algorithm achieves an average TPR of 0.65 and 0.73 with a timing error tolerance region (TETR) of 3 and 4 seconds respectively for the detection of the onsets.

For the features based on the generalized Hurst exponent, our algorithm achieves an average TPR of 0.69 and 0.77 with a TETR of 3 and 4 seconds respectively for the detection of the onsets.

%%%%%%%%%%%%%%%%%%%%%%%%%%%%%%%%%%%%%%%%%%%%%%%%%%%%%%%%%%%%%%%%%%%%%%%%%%%%%%%%%%%%%%%%%%%%%%%%
\section{Methods}\label{sec:methods}

\subsection{Common Average Reference} 
\hspace{\parindent} The Common Average Reference (CAR) method, is a technique used in digital signal processing to obtain a higher signal to noise ratio in signals. It consists on subtracting from each sample the information that is present in all channels, it is defined as

\begin{equation}
    V_i^{CAR} = V_i - \frac{1}{n} \sum\limits_{j=1}^n V_j^{ER},
\end{equation}

where $V_j^{ER}$ is the potential between the $i$-th electrode and the reference, and $n$ is the number of electrodes (channels).

\subsection{Statistical values}
\hspace{\parindent} Eight values are used as features, four of them are statistical values commonly known: Mean ($\mu$), Max, Min, Sum. And four of them are values not commonly known: Skewness ($\tilde{\mu}_3$), Kurtosis ($\tilde{\mu}_4$), Shannon entropy $S(x)$ and generalized Hurt exponent $H(q)$.

\vspace{1em}
\textit{Skewness:} Is a measure of how asymmetrical a distribution around the mean is. For a set $X = \{ x_i, x_2, ..., x_n \}$, skewness is defined as follows
\begin{equation}
    \tilde{\mu}_3 = \frac{\mu_3}{\sigma_3} = \frac{E[(X-\mu)^3]}{(E[(X-\mu)^2])^{3/2}},
\end{equation}
where $\mu_3$ is the third centralized moment and $\sigma$ is the standard deviation.

\vspace{1em}
\textit{Kurtosis:} Is a measure of the shape of a distribution curve around the mean. For a set $X = \{ x_i, x_2, ..., x_n \}$, kurtosis is defined as follows
\begin{equation}
    \tilde{\mu}_4 = \frac{\mu_4}{\sigma_4} = \frac{E[(X-\mu)^4]}{(E[(X-\mu)^2])^{2}},
\end{equation}
where $\mu_4$ is the fourth centralized moment and $\sigma$ is the standard deviation.

\vspace{1em}
\textit{Shannon Entropy:} In information theory, the entropy $S(x)$ is the average rate at which information is produced by a stochastic source of data. The higher the Shannon entropy, the bigger the information is given by a new value in the process. It is also defined as the number of necessary bits to encode the information in the process. For a time series $X = \{ x_i, x_2, ..., x_n \}$, entropy is defined as follows
\begin{equation}
    S(X) = -\sum\limits_{i=1}^N p(x_i) \log_2(p(x_i))
\end{equation}
where $p(x_i)$ is the probability of obtaining the value $x_i$.

\vspace{1em}
\textit{Generalized Hurst Exponent:} The Generalized Hurst Exponent $H(q)$  \cite{DiMatteo2003, DiMatteo2007, DiMatteo2005}, is used in time series analysis and fractal analysis as a measure of scaling properties by the $q^{th}$ order moments of the distribution of the increments. For a time series $X = \{ x_i, x_2, ..., x_n \}$, the generalized Hurst exponent can be obtained from the relations in eq. (5) and eq. (6)
\begin{equation}
    K_q (\tau) \sim \left( \frac{\tau}{\nu} \right)^{q H(q)},
\end{equation}
where
\begin{equation}
    K_q(\tau) =  \frac{\langle | X(t+\tau) - X(t) |^q \rangle}{\langle | X(t) |^q \rangle},
\end{equation}
where $K_q(\tau)$ is given from $X(t)$, with $t= \nu,$ $2\nu, ..., k\nu, T$ (scale observation period $T$, and time resolution $\nu$).

For $q=1$, the generalized Hurst exponent is closely related with the original Hurst exponent, which measures how chaotic or unpredictable is a series. Original Hurst exponent is related with fractal dimensions and has been in the study of seizures in the temporal lobe in animals with EEG \cite{Martinez-Gonzalez2017}.

%%%%%%%%%%%%%%%%%%%%%%%%%%%%%%%%%%%%%%%%%%%%%%%%%%%%%%%%%%%%%%%%%%%%%%%%%%%%%%%%%%%%%%%%%%%%%%
\section{Results}\label{sec:Results}
\hspace{\parindent}The objective of this experiment is to identify the onset of imagined words in continuous EEG signals. The dataset of signals consists of recordings of 27 subjects described on \cite{Torres-Garcia2016}. Each element (subject) has 5 signals recordings, each of them containing 33 repetitions of one of 5 imagined words in Spanish: “arriba”, “abajo”, “izquierda”, “derecha” and “seleccionar”. Which mean: up, down, left, right and select respectively. The process is performed in two stages:

\vspace{1em}
\textit{Training stage:} First, from each subject, 5 Folds are made with the 5 signals by extracting one of them and using it as test signal, and the other four signals are kept for the training stage. This procedure is made 5 times in order to obtain a cross validation scheme (Each fold contains 4 signals for training and 1 signal for testing).

In each of the signals, we know \textit{a priori} the markers of the beginning and ending of each imagined word, these markers are going to help us to collect a training corpus and to compare the predicted onsets of the classifiers with the real onsets to calculate a TPR.

\begin{figure}[H]
    \centering
    \includegraphics[width=.45\textwidth]{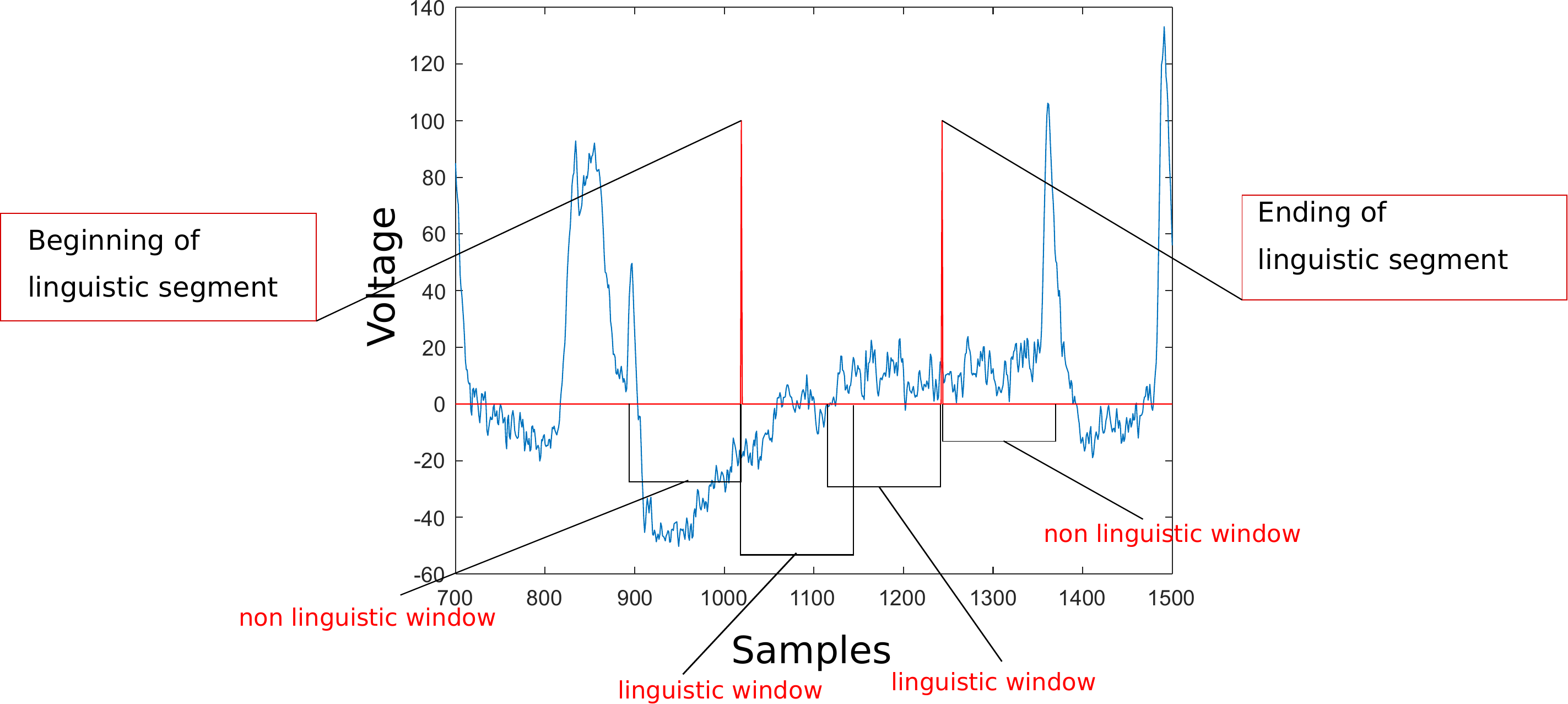}
    \caption{Example of segmentation of signals}
    \label{fig:seg_extr}
\end{figure}

Then, for each fold, from the train signals, two types of segments are selected:

\begin{itemize}
    \item \textbf{Imagined word segments:} From the markers of each repetition of the word, 2 instances are obtained. The first instance is obtained by taking a window of 128 samples right to the marker of beginning of each imagined word. The second segment is obtained by taking a window of 128 samples left to the marker of the ending of each imagined word. From the 33 repetitions, the first and the last are discarded, thus obtaining from the 4 training signals, 248 instances labeled as imagined words.
    \item \textbf{Idle state segments:} Similarly to the imagined word segments, from the markers of each repetition of the word, 2 non-linguistic segments are obtained. The first segment is obtained by taking 128 samples left to the marker of the beginning of each imagined word. The second segment is obtained by taking 128 samples right to the marker of the ending of each imagined word. from the 33 repetitions, the first and the last are also discarded, thus obtaining from the 4 training signals 248 instances labeled as idle states.
\end{itemize}
This process yields 496 instances for each subject, 248 are imagined words and 248 are idle states. This scheme is better illustrated in fig. \ref{fig:seg_extr}.

In each of the 496 instances, noise reduction is performed by applying the CAR method described before and then, two feature sets are obtained. The first feature set is built by calculating 8 values per channel: Mean, Max, Min, Kurtosis, Skewness, Sum of all the samples per channel, Entropy and generalized Hurst exponent with $q=1$. This feature extraction method yields 112 features per instance: 8 features per each channel (14). The second feature set is built by extracting the generalized Hurst exponent for $q=1,2,3,4,5$ of each channel. This feature extraction method yields 70 features per instance. Two Random Forest classifiers are separately trained for each feature set.

\vspace{1em}
\textit{Testing stage:}
With the remaining signal of the fold in each feature set, we test our classifiers by following the steps below.
\begin{enumerate}
    \item The test signal is segmented into windows of 1 second (128 samples), each window is taken sequentially and with no overlap.
    \item Then, each 1 second window is also preprocessed with CAR and the feature extraction method of each feature set is respectively calculated as we did with the training samples, creating with this an instance for each window.
    \item Then the instances are evaluated with the Random Forest classifier of each feature set to define if it is an imagined word or an idle state.
    \item For each classified window, a value is given from the classification: 0 if it is an idle state and 1 if it is an imagined word. The results are appended sequentially into a vector. With this vector, we calculate the onset and ending of each imagined word by unifying all 1's into imagined words and all 0's into idle states, with this, we calculate the predicted onset in time as $n \cdot 128$, being $n$ the position of the first  from an imagined word in the vector of sequential classifications.
    \item The predicted onsets of imagined words are compared with the true onsets to see how far the classification was. This is made by setting up a TETR [3] of 3 seconds (384 samples) and 4 seconds (512 samples) to verify if the onset is inside the TETR window in order to consider it as a True Positive.
\end{enumerate}
This procedure is repeated for the 5 folds and the TPR is averaged..

\subsection*{Results of the onset identification}
\begin{figure}[H]
    \centering
    \includegraphics[width=0.45\textwidth]{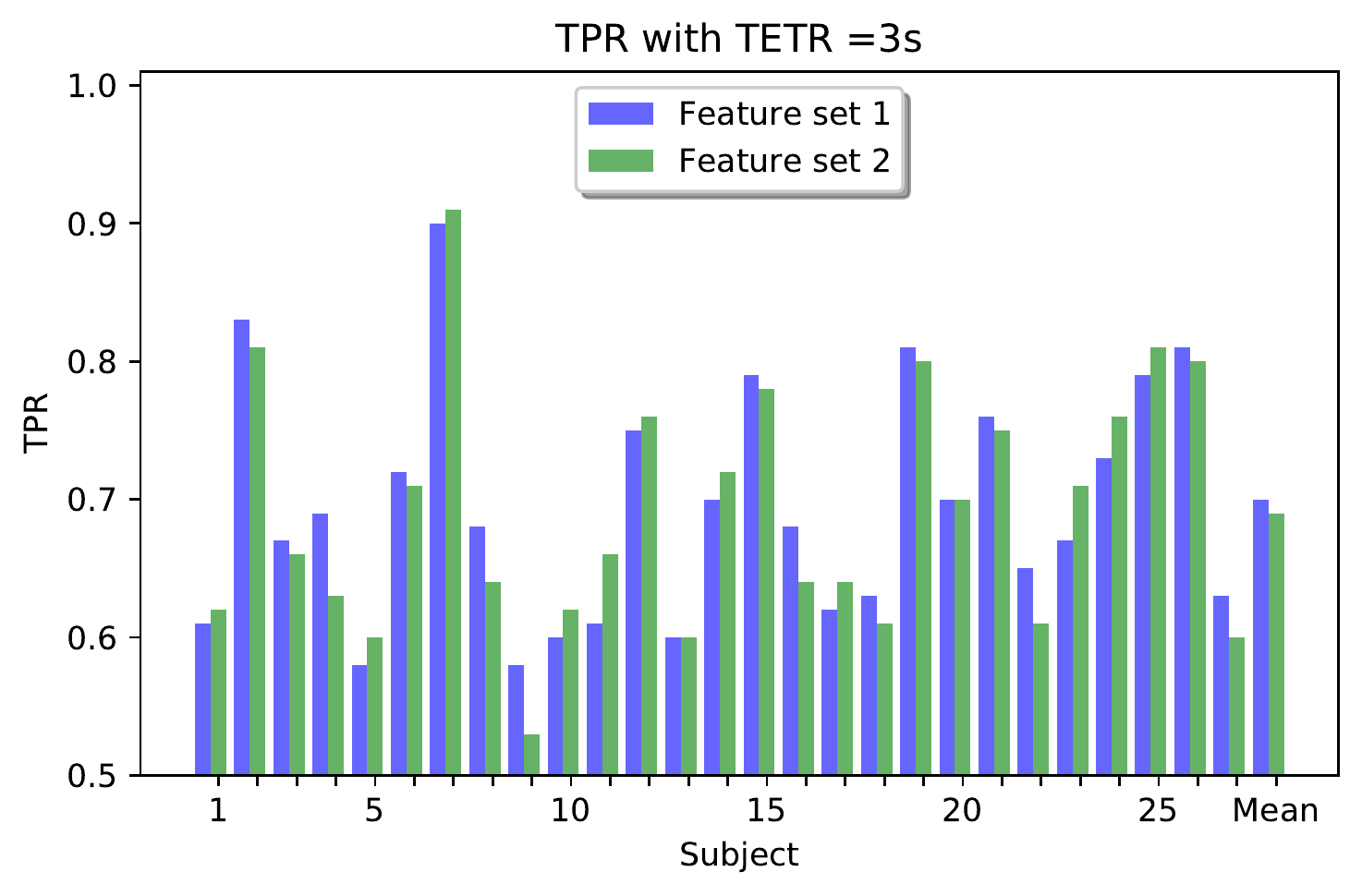}
    \caption{True Positive rate obtained in both feature sets with a TETR of 3 seconds}
    \label{fig:tpr1}
\end{figure}

\begin{figure}[H]
    \centering
    \includegraphics[width=0.45\textwidth]{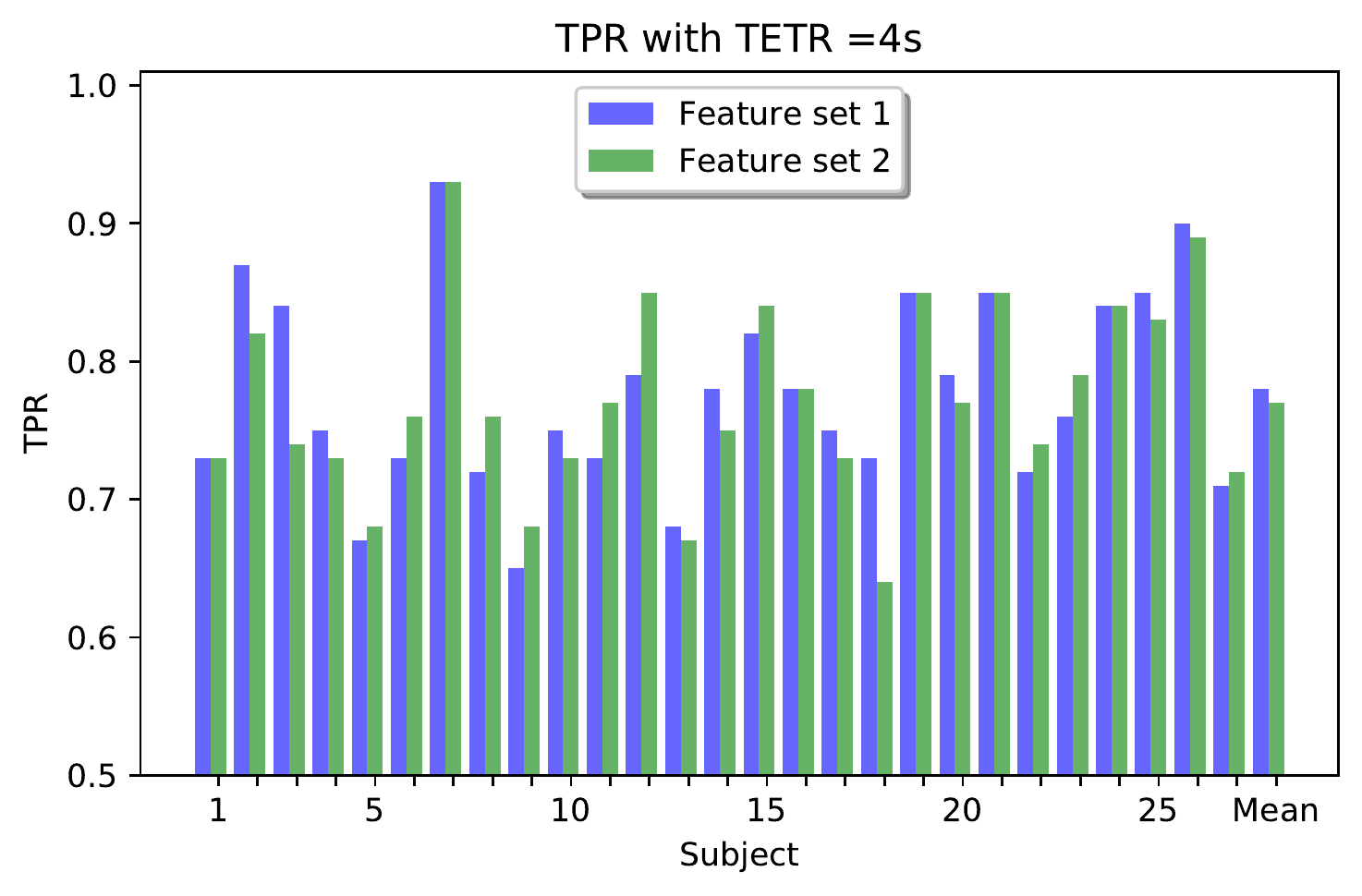}
    \caption{True Positive rate obtained in both feature sets with a TETR of 4 seconds}
    \label{fig:tpr2}
\end{figure}

\hspace{\parindent}The predicted onsets are compared with the real onsets to measure the TPR of onsets, this is, the number of onsets that are correctly identified with a TETR as suggested on \cite{Song2017} for 3 and 4 seconds, then divided by the total number of onsets. Figs. \ref{fig:tpr1} and \ref{fig:tpr2} show the results obtained for each subject on both feature sets.

\section{Discussion}
\hspace{\parindent}From figs. 2 and 3 we can see that in the TETR of 3 seconds, the average TPR on detecting the onsets was 0.65 for the first feature extraction method and 0.69 for the second one. In the TETR of 4 seconds, the average TPR on detecting the onsets was 0.73 for the first feature extraction method and 0.77 for the second one.

Between the two feature extraction methods the highest TPR is obtained by using the second one based only in generalized Hurst exponent, which suggests that the use of this features to identify the onsets of imagined words on continuous signals is more accurate.

The TETR is centered in the real onset which shows us that the error of detecting the onset is in the worst case, half of the TETR (1.5, 2 seconds respectively).

The average span of an imagined word in the corpus is 256 samples (2 seconds). Thus this result looks promising for the future research work. Nevertheless, there is still needed to implement another measure that penalizes the false positives detected by the model. In \cite{Song2017} the True False Positive Rate is introduced which could give a better visualization of this results, thus this new measure will be implemented as future work.

\section{Conclusion}
\hspace{\parindent}In this paper, two feature extraction methods were calculated to test the possibility of detecting the onset of imagined words in EEG signals using a dataset of 27 subjects that imagined 5 different words. The results showed that in the first feature extraction method based on statistical values, the TPR obtained was 0.65 and 0.73 for TETRs of 3 and 4 seconds respectively.

On the second feature set based only in generalized Hurst exponent, the TPR on detecting the onsets of imagined words was 0.69 and 0.77 for TETRs of 3 and 4 seconds respectively.

\section{Additional notes}
The present work is the arXiv version of the work that was presented on the 11th MAVEBA 2019 \cite{Hernandez-Del-Toro2019}. The codes for the work are in the github repository \url{https://github.com/tonahdztoro/onset_detection}.

\section*{References}
\begingroup
\renewcommand{\section}[2]{}
\bibliographystyle{unsrt}
\bibliography{bib}

\begin{thebibliography}{1}

\bibitem{Song2017a}
Youngjae Song and Francisco Sepulveda.
\newblock {An online self-paced brain-computer interface onset detection based
  on sound-production imagery applied to real-life scenarios}.
\newblock In {\em 5th International Winter Conference on Brain-Computer
  Interface, BCI 2017}, pages 46--49, 2017.

\bibitem{Song2017}
Youngjae Song and Francisco Sepulveda.
\newblock {A novel onset detection technique for brain-computer interfaces
  using sound-production related cognitive tasks in simulated-online system}.
\newblock {\em Journal of Neural Engineering}, 14(1), 2017.

\bibitem{Torres-Garcia2016}
Alejandro~A. Torres-Garc{\'{i}}a, Carlos~A. Reyes-Garc{\'{i}}a, Luis
  Villase{\~{n}}or-Pineda, and Gregorio Garc{\'{i}}a-Aguilar.
\newblock {Implementing a fuzzy inference system in a multi-objective EEG
  channel selection model for imagined speech classification}.
\newblock {\em Expert Systems with Applications}, 59:1--12, 2016.

\bibitem{DiMatteo2003}
T.~{Di Matteo}, T.~Aste, and M.~M. Dacorogna.
\newblock {Scaling behaviors in differently developed markets}.
\newblock {\em Physica A: Statistical Mechanics and its Applications},
  324(1-2):183--188, 2003.

\bibitem{DiMatteo2007}
T.~{Di Matteo}.
\newblock {Multi-scaling in finance}.
\newblock {\em Quantitative Finance}, 7(1):21--36, 2007.

\bibitem{DiMatteo2005}
T.~{Di Matteo}, T.~Aste, and Michel~M. Dacorogna.
\newblock {Long-term memories of developed and emerging markets: Using the
  scaling analysis to characterize their stage of development}.
\newblock {\em Journal of Banking and Finance}, 29(4):827--851, 2005.

\bibitem{Martinez-Gonzalez2017}
Claudia~Lizbeth Mart{\'{i}}nez-Gonz{\'{a}}lez, Alexander Balankin, Tessy
  L{\'{o}}pez, Joaqu{\'{i}}n Manjarrez-Marmolejo, and Efra{\'{i}}n~Jos{\'{e}}
  Mart{\'{i}}nez-Ortiz.
\newblock {Evaluation of dynamic scaling of growing interfaces in EEG
  fluctuations of seizures in animal model of temporal lobe epilepsy}.
\newblock {\em Computers in Biology and Medicine}, 88(March):41--49, 2017.

\bibitem{Hernandez-Del-Toro2019}
Tonatiuh Hern{\'{a}}ndez-Del-Toro and Carlos~A. Reyes-Garc{\'{i}}a.
\newblock {An algorithm for detecting the onset of linguistic segments in
  continuous electroencephalogram signals}.
\newblock In {\em 11th Models and Analysis of Vocal Emissions for Biomedical
  Applications (MAVEBA)}, pages 249--252, Florence, Italy., 2019. Firenze
  University Press.

\end{thebibliography}
\endgroup

\end{multicols}
\end{document}